\documentclass{jfm}
\usepackage{graphicx}
\usepackage{caption}
\usepackage{subcaption}
\usepackage{epstopdf, epsfig}
\usepackage{color}
\usepackage{amsmath}

\usepackage[normalem]{ulem}
\usepackage{natbib}

\newcommand{\prob}{p_b}
\newcommand{\weber}{We}
\newcommand{\epsr}{\widetilde{\epsilon}}

\shorttitle{Deformation and breakup of Hinze-scale bubbles}
\shortauthor{A. Masuk, A. Salibindla and R. Ni}

\title{Simultaneous measurements of deforming Hinze-scale bubbles with surrounding turbulence}

\author{Ashik Ullah Mohammad Masuk\aff{1}, Ashwanth K. R. Salibindla\aff{1} \and Rui Ni\aff{1} \corresp{\email{rui.ni@jhu.edu}}}

\affiliation{\aff{1}Department of Mechanical Engineering, Johns Hopkins University, Baltimore, MD 21218, USA}

\begin{document}

\maketitle

\begin{abstract}


We experimentally investigate the breakup mechanisms and probability of Hinze-scale bubbles in turbulence. The Hinze scale is defined as the critical bubble size based on the critical mean Weber number, across which the bubble breakup probability was believed to have an abrupt transition from being dominated by turbulence stresses to being suppressed completely by the surface tension. In this work, to quantify the breakup probability of bubbles with sizes close to the Hinze scale and to examine different breakup mechanisms, both bubbles and their surrounding tracer particles were simultaneously tracked. From the experimental results, two Weber numbers, one calculated from the slip velocity between the two phases and the other one acquired from local velocity gradients, are separated and fitted with models that can be linked back to turbulence characteristics. Moreover, we also provide an empirical model to link bubble deformation to the two Weber numbers by extending the relationship obtained from potential flow theory. The proposed relationship between bubble aspect ratio and the Weber numbers seems to work consistently well for a range of bubble sizes. Furthermore, the time traces of bubble aspect ratio and the two Weber numbers are connected using the linear forced oscillator model. Finally, having access to the distributions of these two Weber numbers provides a unique way to extract the breakup probability of bubbles with sizes close to the Hinze scale. 

\end{abstract}

\begin{keywords}
Authors should not enter keywords on the manuscript, as these must be chosen by the author during the online submission process and will then be added during the typesetting process (see http://journals.cambridge.org/data/\linebreak[3]relatedlink/jfm-\linebreak[3]keywords.pdf for the full list)
\end{keywords}

\section{Introduction}

The process by which finite-sized gas bubbles and liquid droplets break in a turbulent environment constitutes one of the most fundamental and practically important phenomena in multiphase flows. Details of how this takes place have significant impact in various industrial and natural processes, such as chemical reactors \citep{jakobsen2014chemical}, bioreactors \citep{kawase1990mathematical}, air-sea gas transfer \citep{liss1986air}, drag reduction \citep{lohse2018bubble,verschoof2016bubble}, and two-phase heat transfer \citep{lu2008effect,lu2005drag,dabiri2013transition}. Bubbles in strong turbulence can deform, break, and coalesce with each other. The presence of deformation adds to a problem that is already complicated even for the dispersed two-phase flows with rigid, non-deformable particles \citep{balachandar2010turbulent}. Moreover, most works on bubble deformation have been limited to simulations (\citet{elghobashi2019direct} and the references within) with very few experimental works being able to resolve both phases in 3D simultaneously. It is thus the main objective of this paper to overcome this technical challenge and provide new experimental results to study bubble deformation and breakup in turbulence.

The earliest studies on bubble breakup in turbulence were conducted by \citet{kolmogorov1949breakage} and \citet{hinze1955fundamentals}. In particular, Hinze unified the results of numerous preceding investigations. In his seminal work, he argued that only two dimensionless numbers are needed: one is the Weber number $We$ (also used by \citet{kolmogorov1949breakage}) and the other one is the viscosity group, $N=\mu_d/\sqrt{\rho_d\sigma D/2}$, in which $\rho_d$ and $\mu_d$ are the density and the dynamic viscosity of the dispersed phase, respectively. $\sigma$ is the surface tension, and $D$ is the bubble diameter. This is the first time that the idea of the critical Weber number was introduced, and Hinze argued that the critical Weber number must depend only on $N$ following $We_{crit}=c(1+f(N))$, where $f(N)$ is a function of $N$. For bubbles with vanishing inner viscosity, the critical Weber number should just be a constant $c$. A critical Weber number of 0.59 was extrapolated from an earlier experiment conducted by \citet{clay1940mechanism}. In this work, the  Weber number is defined based on external stresses $\tau$ applied on the bubble surface $We=\tau D/\sigma$. $\tau$ is related to the energy dissipation rate ($\epsilon$) in the form of $\tau=C_2(\epsilon D)^{2/3}$ based on the Kolmogorov theory, in which $C_2\approx 2.13$ is the Kolmogorov constant. This formulation should be, strictly speaking, only applied to homogeneous and isotropic turbulence, yet it has been used in many other flow configurations, including chemical reactors with impellers and jets based on the assumption of local isotropy.

Hinze's framework was constructed primarily for liquid droplets. But he noted that the critical Weber number should not be a universal constant; instead, it depends on the density difference between the two phases. \citet{sevik1973splitting} extended this framework to bubbles splitting in turbulence, in which a large density difference between the two phases was present. A slightly larger critical Weber number of 1.26 was observed. By assuming bubbles break once they start to resonate with surrounding turbulent eddies, the critical Weber number can be calculated analytically by equating the natural frequency of bubbles \citep{lamb1932hydrodynamics}  with the reciprocal of the eddy turnover time. The predicted value seems to agree with their measured results. Although $We_{crit}$ has been studied and reported in different types of flows, it should be noted that no consensus on $We_{crit}$ has been reached so far. For air bubble breaking in different flow configurations, such as linear shear flow, turbulent jets, and homogeneous isotropic turbulence, a wide range of $We_{crit}$ from 0.59 to 7.8 have been reported to date \citep{hinze1955fundamentals,sevik1973splitting,deane2002scale,martinez1999breakup1,risso1998oscillations}. Based on this observation, one can only conclude that $We_{crit}$ is roughly of order unity.



Introducing $We_{crit}$ also comes with a critical length scale. For a given mean turbulence energy dissipation rate $\langle\epsilon\rangle$ ($\langle . \rangle$ denotes ensemble average), the critical bubble size is often referred to as the Hinze scale $D_H$, and it is related to $We_{crit}$ following $We_{crit}=\rho C_2(\langle\epsilon\rangle D_H)^{2/3}D_H/\sigma$. For a given $We_{crit}$, it is important to introduce the idea of energy-abundant/super-Hinze ($We\gg We_{crit}$ and $D\gg D_H$) versus energy-limited/sub-Hinze ($We< We_{crit}$ and $D< D_H$) breakups. The former one has been studied much more extensively than the latter for a simple reason: super-Hinze breakup is much faster and more frequent so it is easier to observe in a finite volume and to collect enough statistics. Super-Hinze breakup is typically studied in several different flow configurations: pipe flow \citep{hesketh1991experimental} and turbulent jets \citep{sevik1973splitting,martinez1999breakup,vejravzka2018experiments}. In these cases, the energy contained in turbulent eddies is so abundant that each bubble is almost guaranteed to break--it is only a matter of {\it{time}}.

Breakup frequency is an important parameter in the population balance equation \citep{hulburt1964some,ramkrishna2000population}. However, this framework has one limitation---it assumes that all bubbles above the Hinze scale will eventually break and no bubbles below the scale will ever break. This poses an important challenge to numerical simulations to account for sub-Hinze scale microbubbles, which are important to air-sea gas exchange \citep{deane2002scale}, as well as underwater acoustics as these small bubbles tend to remain in the waterside for an extended period of time.



The breakup mechanisms that have been proposed and evaluated in the literature include (i) persistent stretching by straining flows (parallel flow, plane hyperbolic, axisymmetric hyperbolic, Couette flow, or rotating flow) \citep{hinze1955fundamentals}; bubbles tend to exhibit regular affine deformation in these types of flows (lenticular or cigar-shaped). (ii) resonance mechanism that relies on bubble oscillation to siphon energy from the surrounding turbulence until breakup \citep{sevik1973splitting, hesketh1991experimental,risso1998oscillations}. It typically assumes that the surrounding eddies retain a similar frequency with bubbles' natural frequency. (iii) Inertial mechanism relies on bubbles suddenly being exposed to strong flows, which leads to an almost-immediate irregular breakup. This mechanism has been studied in many contexts in addition to turbulence-induced breakup, e.g. raindrop fragmentation \citep{villermaux2009single} and bag breakup in crossflows \citep{ng2008bag}. In turbulence, three mechanisms may be all present, so applying only one mean Weber number to account for all breakup mechanisms is questionable.

In addition, as \citet{risso1998oscillations} noted, the instantaneous and local Weber number, $We$, could be much larger than the mean value $\langle We\rangle$. They proposed to use the time trace of $We$ along each bubble trajectory to evaluate its breakup frequency. However, in their experiments, such instantaneous Weber number was not directly accessible. As a result, flow velocity from single-phase turbulence was used as a surrogate. This is a common practice in the community as the simultaneous measurements of both phases, either in 2D or 3D, remain  challenging.

To resolve deformation and breakup of the Hinze or sub-Hinze scale bubbles, in this paper, we will introduce an experiment that provides unique simultaneous measurements of both bubble deformation and surrounding flows thanks to the recent advancement of the 3D high-concentration particle shadow tracking \citep{tan2019open} and 3D virtual-camera visual-hull shape reconstruction \citep{masuk2019robust}. In \S\ref{sec:exp}, the experimental setup, i.e. a vertical water tunnel system with a large section of homogeneous and isotropic turbulence, will be introduced. In the same section, the optical system designed to conduct simultaneous measurements of both the phases will also be discussed. In \S\ref{sec:deform}, based on the new datasets, we discuss how flow decomposition can be conducted to analyze the relative roles played by different mechanisms. In \S\ref{sec:breakup_probability}, we finally estimate the breakup probability of Hinze-scale bubbles in turbulence.

\section{Experimental Setup}
\label{sec:exp}

\begin{figure}
    \centering
    \includegraphics[width=0.7\linewidth]{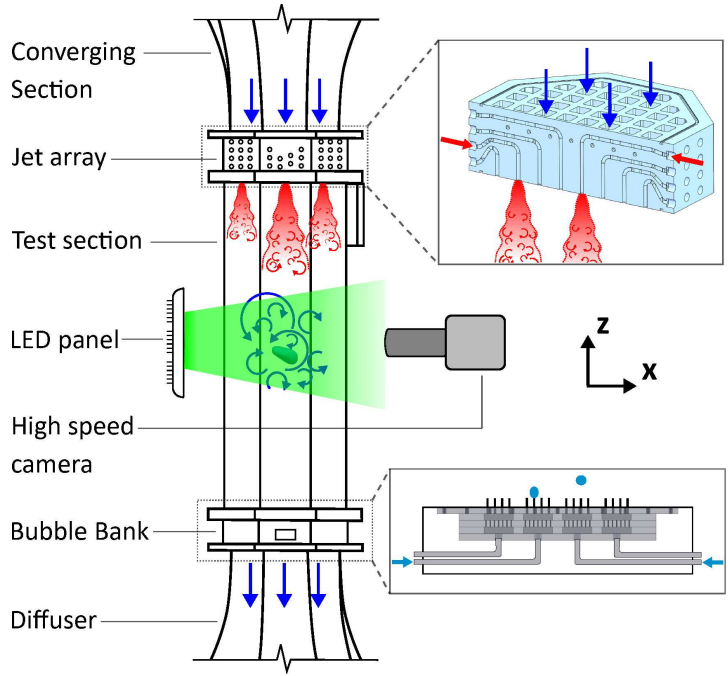}
    \caption{Schematic of the V-ONSET vertical water tunnel; two insets show the 3D model of the jet array used to fire high-speed water jets into the test section and a bubble bank to inject bubbles, respectively. Additional details concerning this facility can be found in \citet{masuk2019v}}
    \label{tunnel}
\end{figure}

A facility was designed to accomplish two main goals: (i) maintain homogeneous and isotropic turbulence in a large volume to ensure that bubbles within this volume experience similar turbulence characteristics, and (ii) bubble deformation should be driven primarily by turbulence rather than by buoyancy, and bubble sizes remain close to the Hinze scale so that we can investigate the deformation and breakup of the Hinze-scale and sub-Hinze-scale bubbles. Satisfying both criteria is challenging. For example, many systems that feature a large region of homogeneous and isotropic turbulence tend to have a low energy dissipation rate \citep{variano2004random, mercado2012lagrangian} ($\langle\epsilon\rangle=O(10^{-5}$--$10^{-3})$ m$^2$/s$^3$), whereas facilities that use water jets to break super-Hinze-scale bubbles can generate a large energy dissipation rate $\langle\epsilon\rangle=O(0.1$--$10^{3})$ m$^2$/s$^3$ at the cost of having strong flow inhomogeneity and anisotropy \citep{martinez1999breakup,vejravzka2018experiments}. 

The experimental setup used for the current study is essentially a vertical water tunnel capable of generating turbulence with $\langle\epsilon\rangle$ roughly at $0.16$ -- $0.5$ m$^2$/s$^3$. To extend the residence time of a Hinze-scale bubble in the interrogation volume, the mean flow in the tunnel was configured to move downward in a vertically-oriented test section. The flow speed was adjusted to balance the rise velocity of bubbles with diameters at around 3 mm to increase the residence time of these bubbles in the view area. Combined with $\langle\epsilon\rangle$ in this region, $\langle We \rangle$ was roughly at 1.19, indicating that most bubbles in the interrogation volume are close to the Hinze scale.


Turbulence in the test section was generated using 88 high-speed water jets (up to 12 m/s), each of which has a diameter $d_j$ of 5 mm, firing co-axially downward into the test section along with the mean flow. The firing pattern of these momentum jets was randomized in a way similar to the work by \citet{variano2004random} in order to ensure that no secondary flow structure would develop in the test section \citep{de1994oscillating,srdic1996generation,variano2004random}. On average, 12.5\% of the jets were kept on at a time as this was found to maximize the turbulence intensity. The test section was set much farther downstream of the jets (about 80$d_j$) to ensure that the jets were well mixed and turbulence becomes homogeneous and isotropic with very little spatial variation. Additional details concerning this setup and its flow characteristics can be found in \citet{masuk2019v}.

Bubbles were generated at the bottom of the test section using two different sizes of hypodermic needles (Small needles: inner diameter (ID) of 160 $\mu$m and outer diameter (OD) of 300 $\mu$m; large needles: ID of 260 $\mu$m and OD of 500 $\mu$m). The range of bubble diameters in the experiment was 2--7 mm, which was set mostly by turbulence generated in our tunnel as large bubbles were broken before entering the interrogation window. Note that the bubble injection was far below the measurement volume to ensure that bubbles entering the measurement volume already lost any memory of the injection.


Both the bubble dynamics and turbulence statistics were collected by using six high-speed cameras each with a 1024$\times$1024 pixel resolution and 4000 frame per second (fps) frame rate. The frame rate was selected to ensure that about 10 frames of images were taken within one Kolmogorov timescale $\tau_\eta=$2.5 ms. These cameras were spatially distributed to cover the entire perimeter of the octagonal test section. Six red LED panels with wavelength at roughly 630 nm were used to provide diffused backlighting to cast shadows of both particles and bubbles onto the imaging planes of all six cameras.


\section{Flow characterization}
\label{sec:flow}

\begin{figure}
    \centering
    \includegraphics[width=0.58\linewidth]{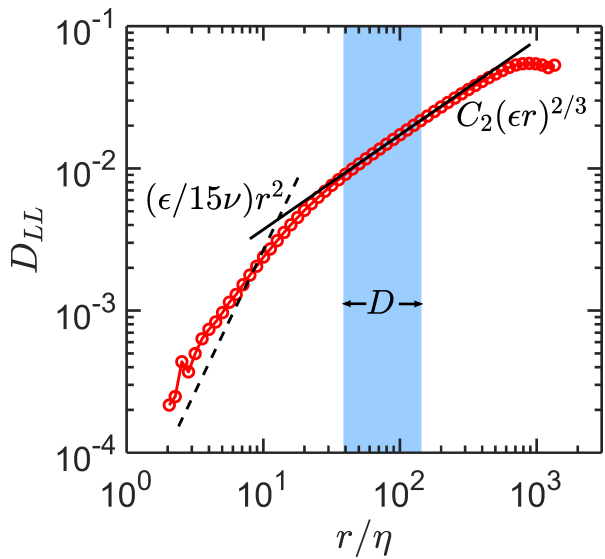}
    \caption{The longitudinal structure function $D_{LL}$ as a function of the scale separation $r$ normalized by the Kolmogorov length scale $\eta$. The dashed and solid lines indicate the dissipative and inertial range scalings based on the Kolmogorov theory, respectively.}
    \label{fig:flow}
\end{figure}

Before discussing bubble deformation and breakup, single-phase turbulence statistics in the tunnel needs to be characterized to ensure that the flow is close to homogeneous and isotropic, and the bubble size is within the inertial range of turbulence. The details of these statistics can be found in \citet{masuk2019v}. Here, we only show the measured longitudinal second-order structure function i.e. $D_{LL}$ in figure \ref{fig:flow}. It can be seen that our experiments were able to resolve length scales as small as 2$\eta$, with $\eta\approx$50 $\mu$m being the Kolmogorov length scale $\eta=(\nu^3/\langle\epsilon\rangle)^{1/4}$. $\nu$ is the kinematic viscosity of water. We can resolve such a small scale thanks to our in-house high-concentration particle tracking system \citep{tan2019open} that employs the  Shake-The-Box method \citep{schanz2016shake}. 

The structure function should approach two limits: one in the dissipative range ($r\ll\eta$) and the other in the inertial range ($\eta\ll r\ll L$). $L$ is the integral scale, which is estimated based on $L\approx u'^3/\langle\epsilon\rangle$, where $u'$ is the fluctuation velocity. The scale separation between $\eta$ and $L$ is determined by the Taylor-scale Reynolds number $Re_{\lambda}=\sqrt{15 u'L/\nu}$, which is estimated to be around 435. In the dissipative range, the structure function follows the relationship of $D_{LL}=(\epsilon/15\nu)r^2$. In the inertial range, the 2/3--scaling law is based on the classical Kolmogorov theory. Although how long the inertial range is and if the Kolmogorov constant $C_2$ is affected by the finite-Reynolds number effect are subjected to further investigations \citep{ni2013kolmogorov}, using a standard number $C_2=2.13$ can provide a reasonable estimation of $\epsilon$. The solid line shown in the figure is based on the calculated $\langle\epsilon\rangle=0.16$ m$^2$/s$^3$. However, if $\langle\epsilon\rangle$ obtained from the inertial range is used to predict the dissipative range $D_{LL}$ (dashed line), it appears that the dashed line is systematically lower than the experimental results. In sum, the difference of  $\langle\epsilon\rangle$ estimated from either the dissipative or inertial range helps to quantify the experimental uncertainty of the mean energy dissipation rate: $\langle\epsilon\rangle$=0.22$\pm$0.07 m$^2$/s$^3$. Moreover, after bubbles getting injected into the system, bubbles can actively modulate turbulence and increase the local energy dissipation rate to around 0.52 m$^2$/s$^3$, which is calculated not from the structure functions but from the local velocity gradients that will be introduced in \S\ref{sec:vg} and figure \ref{fig:epsilon}(b).  

The shaded area in figure \ref{fig:flow}(b) marks the size range of bubbles with respect to the Kolmogorov scale $\eta$. As one can see, bubbles are well within the inertial range of turbulence, indicating that their deformation and breakup are indeed driven by the velocity fluctuations that can be estimated by the inertial range scaling. 


\section{Results and discussions}
\label{sec:deform}

\subsection{Simultaneous bubble and particle tracking}

\begin{figure}
    \centering
    \includegraphics[width=0.7\linewidth]{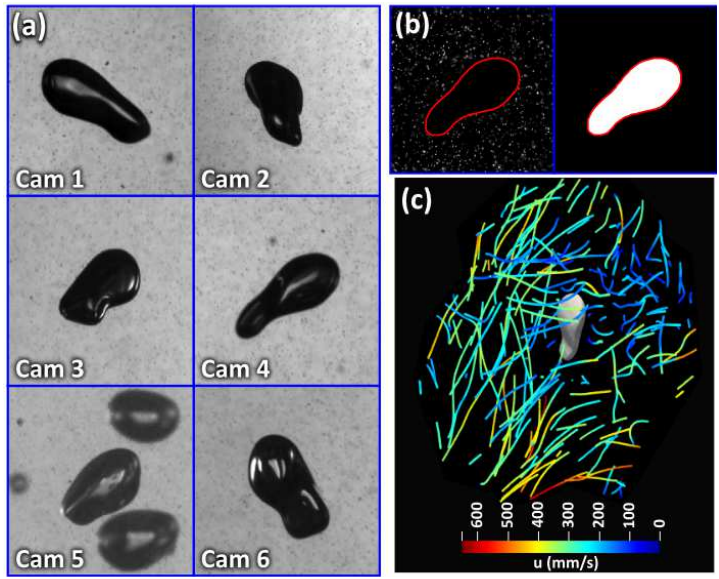}
    \caption{(a) Raw images of one highly-deformed bubble observed by six high-speed cameras simultaneously, (b) The outline and silhouette of the same bubble extracted from Cam 4, (c) 3D tracks of about 40 tracer particles within $4D$ ($D$ is the volume-equivalent sphere diameter) from the center of the bubble that is shown as a 3D reconstructed geometry.}
    \label{fig:images}
\end{figure}

As shown in figure \ref{fig:images}(a), shadows of both bubbles and particles were projected onto the imaging planes of cameras. It is straightforward to separate their images based on the size difference. An example of segmented images of a bubble and surrounding tracer particles is shown in figure \ref{fig:images}(b). The bubble silhouette was then input into a recently-developed virtual-camera visual hull method \citep{masuk2019robust} for 3D shape reconstruction. Averaging surface points on the reconstructed geometry helps to determine the center of mass, which was then tracked in 3D to obtain a bubble trajectory. This procedure was repeated for all bubbles to acquire both the kinematics (from tracks) as well as geometrical information (from 3D shape reconstruction). On average, there were about 15 bubbles in the interrogation volume at each time instant, and each bubble trajectory roughly lasts about 0.09 seconds (360 frames) before exiting. 

Separated images for tracer particles were input into our in-house OpenLPT \citep{tan2019open} to perform the shake-the-box calculation (OpenLPT has already been open-sourced and is available for the entire community to use @JHU-Ni-Lab on Github). Compared with bubbles, significantly more particles could be found in the interrogation volume. At each time instant, there were about 6,000 tracer particles with the mean track length of about 200 frames.

Figure \ref{fig:images}(c) shows one example of about 40 tracer trajectories in the vicinity of a bubble with its 3D shape reconstructed from silhouettes segmented from figure \ref{fig:images}(a). In this case, trajectories of tracer particles within $4D$ away from the bubble center were included. These tracks could be used to estimate the flow condition around the bubble. Since a high-concentration of tracer particles were available around almost every bubble, this experiment provided access to almost all key physical quantities---the Weber number, turbulence energy dissipation rate, and the full coarse-grained velocity gradients---locally, instantaneously, and along each bubble trajectory. Additional information concerning the setup and measurement techniques can be found in works by \citet{masuk2019robust}, \citet{masuk2019v}, and \citet{tan2019open}.


\subsection{Flow velocity and velocity gradient}
\label{sec:vg}

For a bubble at location $\mathbf{x_0}$, its surrounding flow velocity $\mathbf{u}^p(\mathbf{x_0}+\mathbf{x}^p)$ can be measured at a number of discrete positions $\mathbf{x_0}+\mathbf{x}^p$ where $n$ tracer particles are located ($p=1,2,...,n$). These tracer particles are sought within a radius of $D_s/2$ from the bubble center with $D_s$ being the diameter of a spherical search volume. The flow field within this range can be decomposed into leading terms by applying the Taylor expansion:

\begin{align}\label{eqn.vel_decomp}
\begin{split}
    u^p_i(\mathbf{x_0}+\mathbf{x}^p) \approx \overline{u_i}(\mathbf{x_0}) + \widetilde{A}_{ij}(\mathbf{x_0})x^p_j + O\left(x^p_j \widetilde{H}_{jik}(\mathbf{x_0})x^p_k\right)\\
    \widetilde{A}_{ij}(\mathbf{x_0})=\frac{\partial u^p_i}{\partial x^p_j}~~~~\text{and}~~~~~\widetilde{H}_{jik}(\mathbf{x_0})=\frac{\partial^2 u^p_i}{\partial x^p_j\partial x^p_k}
\end{split}
\end{align}

where $\overline{u_i}=\sum^{n}_{p=1} u^p_i(\mathbf{x_0}+\mathbf{x}^p)/n$ represents the local mean flow estimated by averaging the velocity of $n$ tracer particles. $\widetilde{A}_{ij}$ and $\widetilde{H}_{jik}$ indicate the velocity gradient tensor and the Hessian matrix, respectively, and the tilde denotes coarse-graining at the bubble size. $\mathbf{x}^p$ is the separation vector directed from the bubble center at $\mathbf{x_0}$ to the $p_{th}$ tracer particle location. For small micro-bubbles with sizes in the dissipative range ($D\ll\eta$), the flow is linear so the velocity Hessian is negligibly small. This higher-order term grows as a function of bubble size and eventually becomes important for bubbles with sizes in the inertial range ($\eta\ll D\ll L$).

Although the velocity gradient around each bubble can be measured accurately \citep{ni2015measurements}, the velocity Hessian, on the other hand, requires measuring the gradient of the velocity gradient (three $3\times3$ matrices). Even though it is possible to calculate the velocity Hessian given sufficient number of tracer particles, the uncertainty becomes large due to the second-order spatial derivative. As a result, we limit only to the first two orders, i.e. the mean flow velocity $\overline{u_i}$ and the velocity gradient  $\widetilde{A}_{ij}(\mathbf{x_0})$, to capture the key mechanisms of deformation. 

The velocity gradient tensor $\widetilde{A}_{ij}$ can be uniquely solved if we have four particles around a bubble. In practice, on average, 30--40 particles were used to perform least-squares fit by seeking the minimum value of the squared residuals $\sum_p[u^p_i-\widetilde{A}_{ij}x^p_j]^2$ \citep{pumir2013tetrahedron, ni2015measurements}. Although finite-sized bubbles typically come with a large search radius and abundant nearby tracer particles thanks to our tracking method \citep{tan2020introducing}, particles in the vicinity of a bubble are still randomly distributed in space. If nearby particles stay primarily within a quasi-2D plane, the estimation of the out-of-plane velocity gradient will have large uncertainty. Similar to previous studies \citep{xu2011pirouette,ni2015measurements}, an inertia tensor $I=\sum_px^p_ix^p_j/tr(\sum_px^p_ix^p_j)$ was adopted to evaluate the shape factor of the particle cloud. If particles are uniformly distributed in 3D, three eigenvalues of this inertia tensor ($\gamma_i$) all equal to 1/3. For a quasi-2D distribution, the smallest eigenvalue ($\gamma_3$) will be very close to zero, and the gradient along that direction cannot be calculated. In practice, events with $\gamma_3/\gamma_1$ smaller than 0.15 was therefore removed from the statistics. 


\begin{figure}
    \centering
    \includegraphics[width=0.95\linewidth]{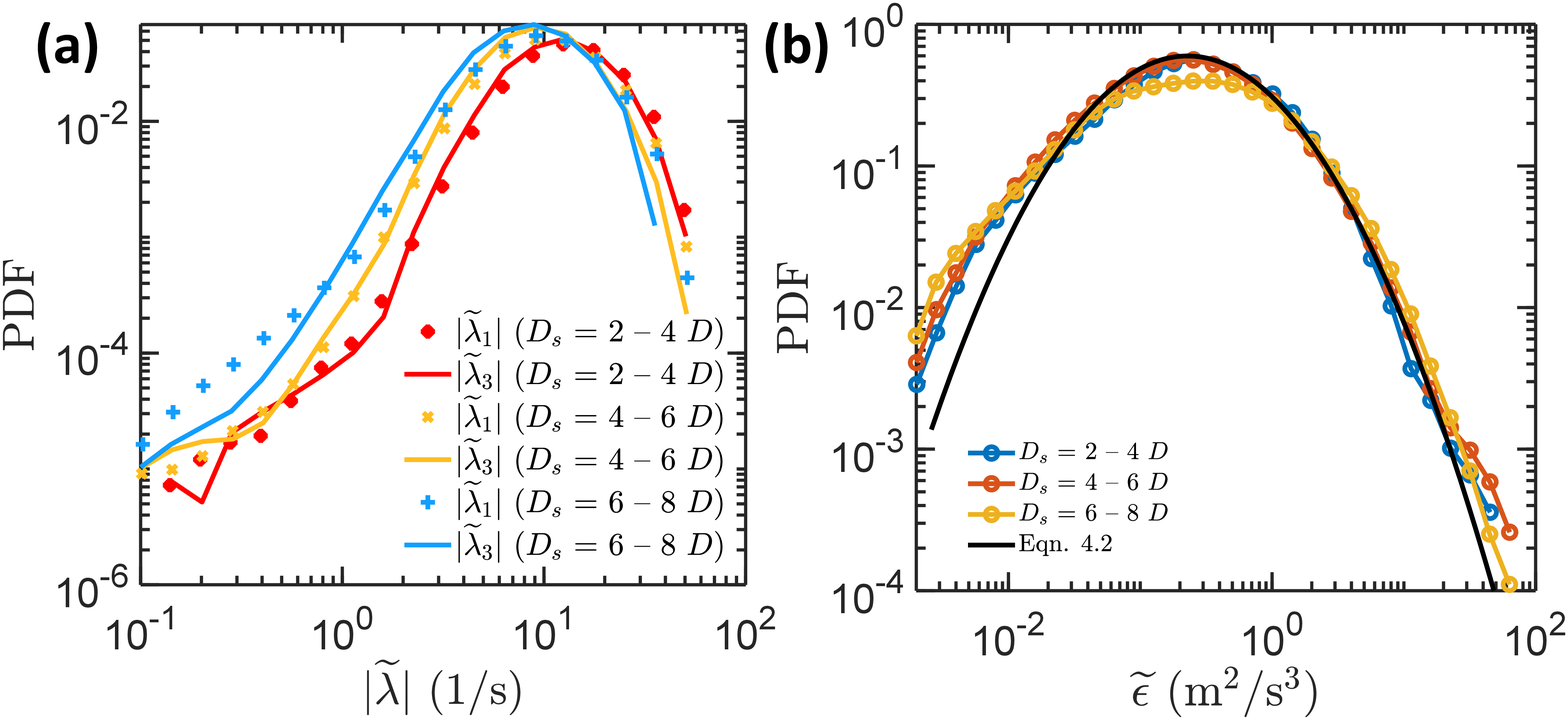}
    \caption{(a) The distribution of the two eigenvalues ($\widetilde{\lambda_1}$ and $\widetilde{\lambda_3}$) of the local rate-of-strain tensor coarsed-grained at the bubble scale $D$ ($|\widetilde{\lambda}|$ is used here because $\widetilde{\lambda_3}<0$.); Three search diameters ranging from from 2--4$D$ to 6--8$D$ are denoted by different colors. (b) The distribution of the local coarsed-grained energy dissipation rate $\epsr$. The log-normal distribution from equation \ref{eq:intm} is shown as the black solid line.}
    \label{fig:epsilon}
\end{figure}


Based on $\widetilde{A}_{ij}$, the coarse-grained rate-of-strain tensor, $\widetilde{S}_{ij}$, and rotation tensor, $\widetilde{\Omega}_{ij}$ can be directly obtained: $\widetilde{S}_{ij}=\frac{1}{2}(\widetilde{A}_{ij}+\widetilde{A}_{ji})$, $\widetilde{\Omega}_{ij}=\frac{1}{2}(\widetilde{A}_{ij}-\widetilde{A}_{ji})$. Figure \ref{fig:epsilon}(a) shows the probability density function (PDF) of two eigenvalues of $\widetilde{S}_{ij}$ (the largest $\widetilde{\lambda_1}$ and the smallest $\widetilde{\lambda_3}$) based on different $D_s$. The PDFs of $|\widetilde{\lambda_1}|$ and $|\widetilde{\lambda_3}|$ overlap with each other for three $D_s$ considered, indicating that the magnitude of flow stretching and compression near a bubble on average is similar. The PDF progressively shifts leftward as $D_s$ becomes larger because coarse-graining velocity gradients at a larger $D_s$ works effectively as enlarging a low-pass filter, which will continue to reduce the gradient as $D_s$ increases. As a result, the calculated velocity gradient using particles within a search diameter of $D_s$ should always underestimate $\widetilde{A}_{ij}$ at the bubble scale $D$ because $D_s>D$. Fortunately, both $D_s$ and $D$ are in the inertial range, and the eigenvalues of $\widetilde{A}_{ij}$ can be related to the local energy dissipation rate in the form of $C_2(\epsr d)^{2/3}=(\widetilde{\lambda_3}d)^2$, where $C_2=2.13$ is the Kolmogorov constant \citep{batchelor1953theory,sreenivasan1995universality, ni2013kolmogorov} and $\epsr$ is the coarse-grained energy dissipation rate for a range of length scales $d$  considered. To check if the distribution of $\epsr$ is indeed the same for $d$ varying between $D$ to $D_s$, in figure \ref{fig:epsilon}(b), the PDFs of the estimated local $\epsr$ using three different $D_s$ are shown. Although the distribution of the calculated  $\widetilde{\lambda_3}$ are sensitive to $D_s$, once converted to $\epsr$, three curves from all three $D_s$ fall right on top of each other, indicating that the local $\epsr$ is roughly the same for the range of $D_s$ considered. Therefore, although $D_s>D$ is needed to include enough tracer particles for calculating velocity gradients, the statistics reported are insensitive to $D_s$ thanks to the universal inertial range scaling in homogeneous and isotropic turbulence. 


The coarse-grained energy dissipation rate can be described by the log-normal distribution based on the Kolmogorov refined theory in 1962 \citep{kolmogorov1962refinement} and multi-fractal spectrum \citep{meneveau1991multifractal}.



\begin{equation}
\label{eq:intm}
    P\left(\frac{\epsilon_r}{\langle\epsilon\rangle}\right) = \frac{1}{\epsilon_r/\langle\epsilon\rangle}\frac{1}{\sqrt{2\pi (A+\mu \ln(L/r))}} \exp\left[-\frac{(\ln(\epsilon_r/\langle\epsilon\rangle) + 1/2(A+\mu \ln(L/r)))^2}{2(A+\mu \ln(L/r))}\right]
\end{equation}

where $\epsilon_r$ is the energy dissipation rate coarse-grained at a scale $r$. $\mu\approx0.25$ is the intermittency exponent. $L=3.2$--6 cm is the integral length scale. $A$ is a parameter that needs to be fitted to the experimental data to determine the variance of $\epsr$ when $r=L$, which was found to be around one. Based on the definition, $\widetilde{\epsilon}$ measured from our experiments is equivalent to $\epsilon_r|_{r=D}$, which is shown as the black solid line in figure \ref{fig:epsilon}. The nice agreement between the experimental data and the log-normal distribution (equation \ref{eq:intm}) shows that the measured coarse-grained energy dissipation rate is consistent with the classical Kolmogorov theory. 


\subsection{Different types of deformation}
\label{sec:we}

\subsubsection{Bubble deformation by the velocity gradient $We_{vg}$}

\begin{figure}
    \centering
    \includegraphics[width=0.6\linewidth]{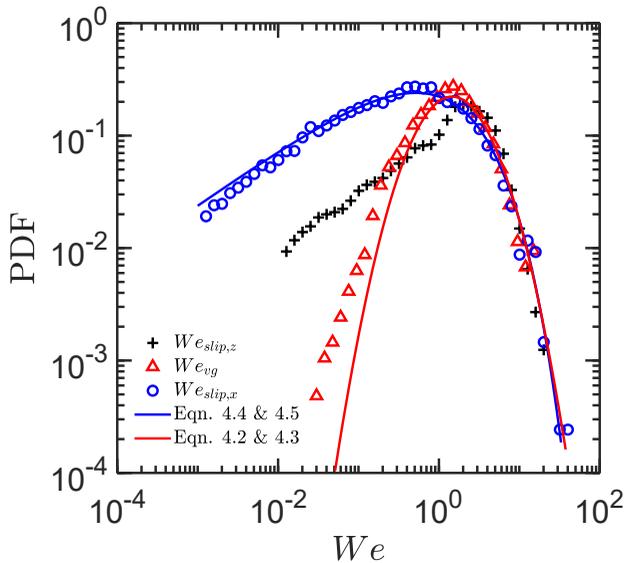}
    \caption{The distribution of the measured Weber numbers, based on the slip velocity, i.e. $We_{slip,x}$ (blue circle) and $We_{slip,z}$ (black plus), and the velocity gradient, $We_{vg}$ (red triangle). Two solid lines represent the modelled Weber number distributions based on the log-normal distribution of $\epsr$ (equation \ref{eq:intm} \& \ref{eq:We_def}, red line) and the stretched-exponential fit of the slip velocity (equation \ref{eq:slip} \& \ref{eq:We_slip}, blue line), respectively.}
    \label{We_comparison}
\end{figure}


In turbulence, the difference of dynamic pressure across a bubble acts to push the bubble interface inward to drive bubble deformation. Based on this argument, $\widetilde{\lambda_3}$, which is associated with the direction that compresses the most, should be the more relevant eigenvalue of $\widetilde{A}_{ij}$. Following the argument, the Weber number can be defined as
\begin{equation}
\label{eq:We_def}
We_{vg}=\frac{\rho(\widetilde{\lambda_3}D)^2D}{\sigma}\sim\frac{C_2(\epsr D)^{2/3}}{\sigma}
\end{equation}


This Weber number definition is based on the local coarse-grained $\widetilde{A}_{ij}$ and $\epsr$, which is different from the mean Weber number defined by \citet{kolmogorov1949breakage} and \citet{hinze1955fundamentals}. Figure \ref{We_comparison} shows the distribution of local $We_{vg}$ based on the measurements of $\widetilde{A}_{ij}$ along each bubble track. The distribution peaks at around one, suggesting that those bubbles are indeed Hinze-scale bubbles. All data points on the right side of the peak represent bubbles deforming under strong velocity gradients. On top of the experimental results, the model of $We_{vg}$ based on equation \ref{eq:intm} and \ref{eq:We_def} is also shown. Similar to figure \ref{fig:epsilon}(b), the log-normal distribution of the local $\epsr$ explains the observed shape of the PDF of $We_{vg}$, from which bubble breakup probability can be determined.




\subsubsection{Slip-velocity induced deformation $We_{slip}$}

As Hinze stated in his original seminal work \citep{hinze1955fundamentals}, employing the velocity gradient to evaluate the deformation and breakup of droplets should only be applied if there is no large density difference between the dispersed phase and the carrier phase. For bubbles in water, such a large density difference does exist, and it is not surprising that $We_{vg}$ may not capture the total stress acted on bubbles by turbulence. For example, the instantaneous velocity mismatch between the two phases could also lead to significant dynamic pressure that needs to be evaluated. This effect can be captured by the so-called slip velocity, $\mathbf{u}_{slip} = \mathbf{u}_b - \mathbf{u}_f$. As its name suggests, $\mathbf{u}_{slip}$ quantifies the drift of a bubble of velocity $\mathbf{u}_b$ away from the instantaneous local flow velocity $\mathbf{u}_f$.

\begin{figure}
    \centering
    \includegraphics[width=0.95\linewidth]{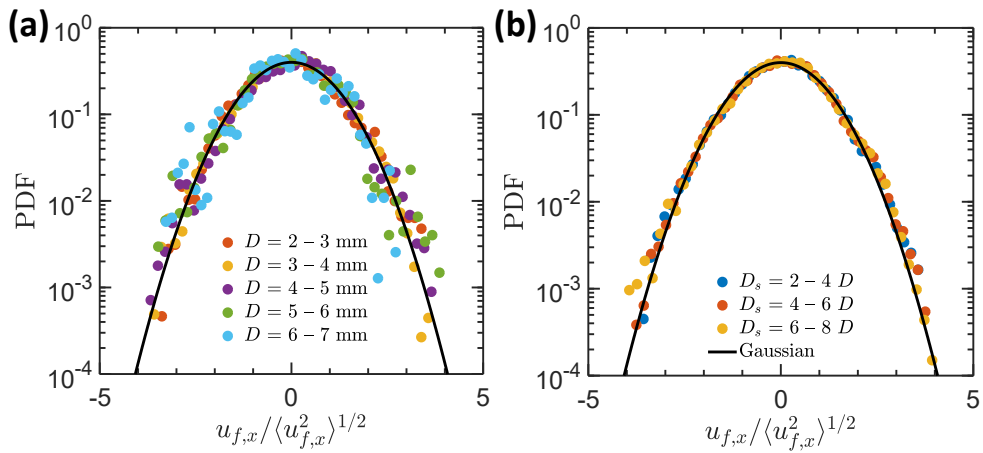}
    \caption{The distribution of the horizontal flow velocity $u_{f,x}$ (normalized by its own standard deviation) nearby bubbles of (a) different sizes $D$, and (b) different search diameters $D_s$. The black solid lines in (a) and (b) show the standard normal distribution for reference.}
    \label{fig:uf_SD}
\end{figure}

\begin{figure}
    \centering
    \includegraphics[width=0.95\linewidth]{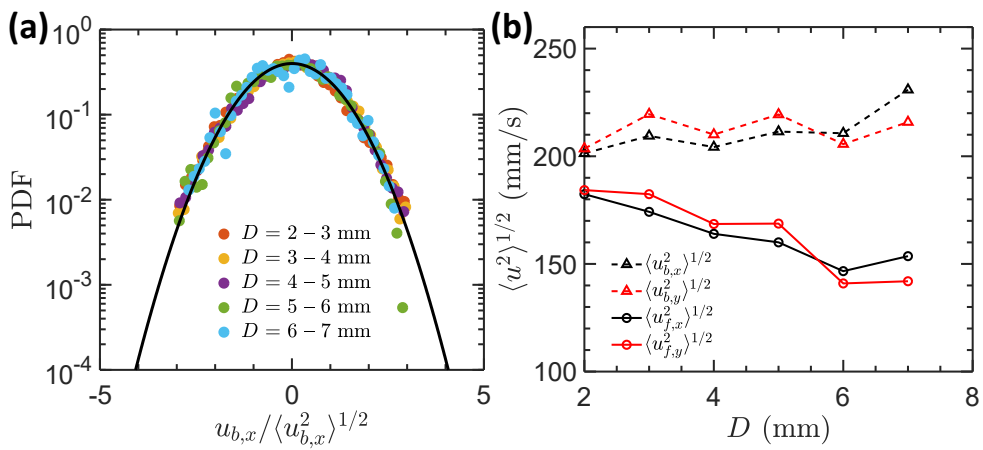}
    \caption{(a) The distribution of the horizontal velocity $u_{b,x}$ of bubbles (normalized by its own standard deviation) with different diameter $D$; The black solid line indicates the standard normal distribution for comparison. (b) The fluctuation of mean flow velocity (solid) and bubble velocity (dashed lines) along two different directions versus bubble size $D$. }
    \label{fig:ub}
\end{figure}

$\mathbf{u}_f$ represents the continuous-phase fluid velocity at the center of a bubble if the bubble was not there. In practice, $\mathbf{u}_f$ needs to be estimated from the continuous-phase velocities measured in the vicinity of the bubble. Therefore, we assume $\mathbf{u}_f$ to be the same as $\overline{u_i}$ in equation \ref{eqn.vel_decomp}, which can be estimated by averaging the tracer velocities around the bubble. Figure \ref{fig:uf_SD}(a) shows the PDF of only one horizontal component of $\mathbf{u}_f$ normalized by its own standard deviation. $\mathbf{u}_f$ can be calculated around bubbles of different sizes, which are shown by different colored symbols. The solid line indicates the standard normal distribution, which seems to agree well with the horizontal velocity distribution of bubbles of all sizes, at least for the range of bubble sizes considered. Since the PDFs of bubbles of all sizes are nearly the same, they can be combined and the results are shown in figure \ref{fig:uf_SD}(b). Furthermore, to rule out the possible $D_s$ effect, the same procedure was repeated for three different $D_s=$2--4$D$ to 6--8$D$. As shown in figure \ref{fig:uf_SD}(b), no discernible difference is observed for the flow velocity PDF at three $D_s$, which suggests that $\mathbf{u}_f$ is not sensitive to $D_s$ either. 



$\mathbf{u}_b$ denotes the bubble velocity with one of its horizontal components along the $x$-axis being  $u_{b,x}$. Figure \ref{fig:ub}(a) shows the distribution of $u_{b,x}$, normalized by its own standard deviation, for a wide range of bubble sizes, and the distribution for all bubble sizes seem to agree with a Gaussian distribution (solid line) very well. The standard deviation of $\mathbf{u}_{b}$ for both horizontal directions are shown as dashed lines in figure \ref{fig:ub}(b), and they exhibit a weak, if at all, dependence on $D$. Note that, in the other limit for bubbles rising in a quiescent medium with no turbulence, since the horizontal velocity is coupled with the size-dependent rise velocity \citep{ern2012wake}, $\langle u_{b,x}^2\rangle^{1/2}$ should depend on the bubble size. Therefore, the observed nearly-constant $\langle u_{b,x}^2\rangle^{1/2}$ clearly indicates that the buoyancy effect is negligible in the horizontal directions due to the background intense turbulence. Figure \ref{fig:ub}(b) also displays the standard deviation of $\mathbf{u}_f$ along two horizontal directions. In contrast to $\langle u_{b,x}^2\rangle^{1/2}$ , $\langle u_{f,x}^2\rangle^{1/2}$ seems to decrease as $D$ increases because a finite-sized bubble effectively serves as a filter that reduces the local flow fluctuations. By extrapolating both $\langle u_{b,x}^2\rangle^{1/2}$ and $\langle u_{f,x}^2\rangle^{1/2}$ to small bubble sizes, $\langle u_{b,x}^2\rangle^{1/2}$ and $\langle u_{f,x}^2\rangle^{1/2}$ will eventually cross at around 200 mm/s for bubble size close to zero, which gives the right limit as extremely-small bubbles should behave similarly to tracers $\langle u_{b,x}^2\rangle^{1/2}\approx\langle u_{f,x}^2\rangle^{1/2}$.

\begin{figure}
    \centering
    \includegraphics[width=0.99\linewidth]{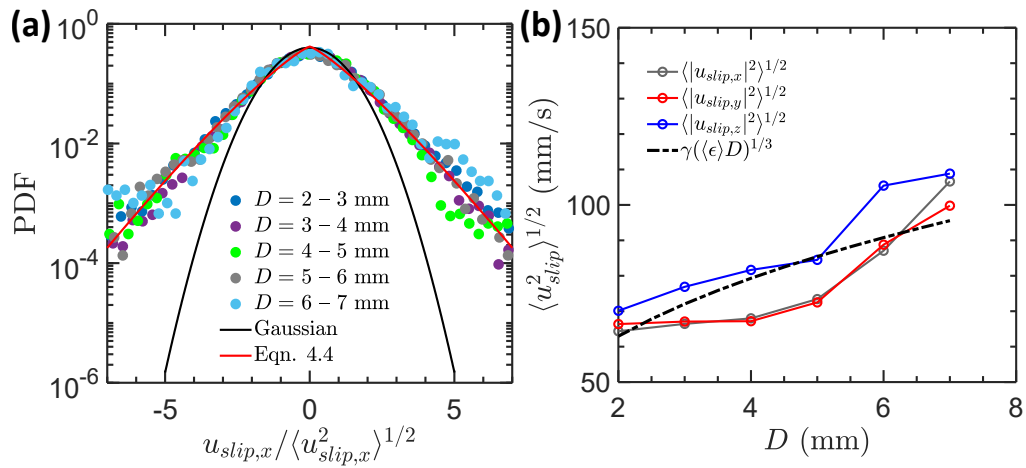}
    \caption{(a) The distribution of the normalized horizontal slip velocity between the two phases; Symbols denote bubbles of different sizes and the black solid line indicates the standard normal distribution. The red solid line shows the stretched exponential (equation \ref{eq:slip}) fit to the data. (b) The fluctuation slip velocity of all three components versus the bubble diameter $D$; The black dash-dotted line indicates the estimation from the second-order structure function. The prefactor $4/9$ is chosen to minimize the offset between the solid line and the data. }
    \label{fig:pdf_uslip}
\end{figure}

Although both $u_f$ and $u_b$ along the horizontal directions appear to follow the Gaussian distribution, the slip velocity $u_{slip}=u_f-u_b$ does not. As shown in figure \ref{fig:pdf_uslip}(a), for bubbles of all sizes, the tails of the slip-velocity PDF ($u_{slip,x}$) are systematically higher than that of the Gaussian function (black solid line), indicating that the slip velocity is more intermittent than the velocity of either phase alone. For the distribution of the normalized slip velocity, similar to the PDFs of $u_{f,x}$ and $u_{b,x}$, no obvious bubble-size dependence is observed. Note that the PDF of $u_{slip}$ resembles that of the velocity increment between two points in single-phase turbulence \citep{kailasnath1992probability, sreenivasan1999fluid, li2005origin}. The PDF of the velocity increment has been fitted with a stretched exponential function \citep{kailasnath1992probability}, which is adopted here to describe the observed PDF of the slip velocity.

\begin{equation}
P(u_{slip,x})=C \exp\left[-Q\left( \frac{u_{slip,x}}{\langle u^2_{slip,x} \rangle^{1/2}}\right)^{m}\right]
\label{eq:slip}
\end{equation}

\begin{equation}
    We_{slip} = \frac{\rho u_{slip}^2 D}{\sigma}
    \label{eq:We_slip}
\end{equation}


where $C$ is the normalization factor, and $Q$ and $m$ are fitting parameters in the stretched exponential function. For single-phase turbulence, the degree to which the tail of the PDF is stretched depends on the scale separation. If the velocity separation is close to the integral length scale, the PDF recovers the Gaussian distribution ($m$=2). As the separation becomes smaller and smaller, the PDF becomes more and more intermittent; at $m=1$, the PDF follows an exponential function. If we take the bubble size 0.03$L$ to 0.12$L$ as the scale separations to calculate the velocity increment in single-phase turbulence, the scaling exponent $m$ should vary between 0.8 to 1.05 based on the work by \citet{kailasnath1992probability}. In our case, although the slip velocity distribution also follows the stretched exponential, the PDF preserves its shape for all bubble sizes considered in this work with no obvious scale dependence, and all symbols in figure \ref{fig:pdf_uslip}(a) collapse with one another. Therefore, the distributions of the normalized slip velocity for different sizes of bubbles were fitted together with one stretched exponential function and one set of constants, i.e. $Q$, and $m$. In particular, $m$ is found to be a constant close to 6/5, which is slightly larger than the range of $m$ from 0.8 to 1.05 in single-phase turbulence. This observation suggests that the slip velocity between the two phases is less intermittent compared with the velocity increment between two points in single-phase turbulence under the same scale separation, which is not surprising since bubbles are capable of filtering out intermittent small-scale fluctuations.


Furthermore, the fluctuation slip velocity ($\langle u_{slip}^2\rangle^{1/2}$) increases as a function of bubble size $D$, suggesting that larger bubbles with a larger inertia tend to deviate further away from the surrounding fluid velocity. At the same time, the typical velocity scale of an eddy of the bubble size $D$ also increases with $D$, following $(\langle\epsilon\rangle D)^{1/3}$. After assuming that these two velocity scales are related, the measured $\langle u_{slip}^2\rangle^{1/2}$ along all three directions are fitted with $\gamma(\langle\epsilon\rangle D)^{1/3}$ by performing the least-square regression to obtain the fitting coefficient $\gamma$, which turns out to be 0.62. The fitted result is shown in figure \ref{fig:pdf_uslip}(b) as black dash-dotted line. It is clear that the fit reproduces the growth of the measured standard deviation of the slip velocity as a function of $D$, but the agreement between the fitted and the measured results is not perfect. Nevertheless, for simplicity and without any other  alternative velocity scales, this fit using the eddy velocity is used to estimate $\langle u_{slip}^2\rangle^{1/2}$ for bubbles with size in the inertial range. With this relationship and two coefficients, i.e. $Q=3/4$, $m=6/5$, the distribution of $u_{slip}$ can be estimated from equation \ref{eq:slip}.



Finally, the distribution of $We_{slip}$, calculated based on equation \ref{eq:slip} and \ref{eq:We_slip}, is shown in figure \ref{We_comparison}. The blue solid line indicates the predicted $We_{slip}$ based on the stretched exponential fit to the horizontal slip velocity $u_{slip,x}$ (equation \ref{eq:slip}). The distribution also peaks at around $We\approx 1$, which is slightly smaller than the most probable value of $We_{vg}$. The right tails of both PDFs ($We_{vg}$ and $We_{slip,x}$), corresponding to the range of $We$ that is important for deformation and breakup, are very close to each other. This may suggest that, for bubble deformation, slip velocity and velocity gradient may be equally important; completely relying on the velocity gradient may not account for all stresses that bubbles experience in turbulence.  

\subsubsection{Buoyancy-induced deformation}

Although the turbulence energy dissipation rate has been set as high as possible in our facility, the buoyancy effect is not negligible. In figure \ref{We_comparison}, the PDF of $We_{slip}$ in the vertical direction based on the $z$-axis slip velocity, i.e. $We_{slip,z}$ is also shown. This PDF has a bump near $We_{slip,z}\approx$3--4 because of the buoyancy effect, but both the left and right tails seem to agree with those of $We_{slip,x}$. This suggests that both the turbulence effect and the buoyancy effect are present in $We_{slip,z}$, but the effect of buoyancy is rather limited to a comparatively narrower region near the peak of the PDF. Nevertheless, the exact functional form of the PDF close to the peak is unknown and requires further investigations to understand the coupling between the local bubble rise velocity and the surrounding turbulence.

Note that $We_{slip,z}$ is similar to the E\"otv\"os number: $Eo=\rho gD^2/\sigma$, as the terminal vertical slip velocity $u_{slip,z}$ driven primarily by buoyancy should be proportional  to $\sqrt{gD}$. Note that this relationship is approximate, as the buoyancy-driven terminal rise velocity is also sensitive to the bubble geometry, orientation, and the drag coefficient. In intense turbulence, these parameters could also be functions of $\epsilon$. In a recent paper \citep{salibindla2020}, the drag coefficient of bubble with different sizes in intense turbulence was reported, and it follows $C_D = {\text{max}}(24/Re_b(1+0.15Re_b^{0.687}),{\text{min}}(f(Eo), f(Eo)/We^{1/3}))$ where $f(Eo) = 8Eo/3(Eo+4)$. Based on this equation, the most probable slip velocity in the vertical direction can be calculated following: $u^2_{slip,z} = 2 V_b (\rho - \rho_b) g / \rho A C_D$ where $V_b$ is the volume of a bubble, $A$ is the projected area of volume-equivalent spherical bubble, and $\rho_b$ is the density of bubble. For the bubble size range considered, $We_{slip,z}$ calculated based on $C_D$ is about 4, which is consistent with the bump of $We_{slip,z}$ observed in the PDF. This agreement confirms that the observed bump in the distribution of $We_{slip,z}$ is indeed driven by buoyancy. 

All together, it seems that the bump in the distribution of $We_{slip,z}$ is limited to a narrow range, and the right tail of $We_{slip,z}$ seems to be close to that of $We_{slip,x}$ and $We_{vg}$. This suggests that, at least for our parameters when $\langle\epsilon\rangle\approx 0.2$--0.5 m$^2$/s$^3$, the buoyancy-induced deformation is limited. If we keep increasing $\langle\epsilon\rangle$, the buoyancy effect will become even weaker.

\subsection{Bubble aspect ratio vs. Weber numbers}

So far, we have focused primarily on discussing the distribution of different definitions of Weber numbers and understanding the connection between these Weber numbers and their associated turbulence characteristics. In this section, the instantaneous Weber numbers along bubble trajectories will be used to study the mechanisms of bubble deformation and breakup in turbulence. 


\subsubsection{Simultaneous measurements of bubble geometry and $We$}

\begin{figure}
    \centering
    \includegraphics[width=0.98\linewidth]{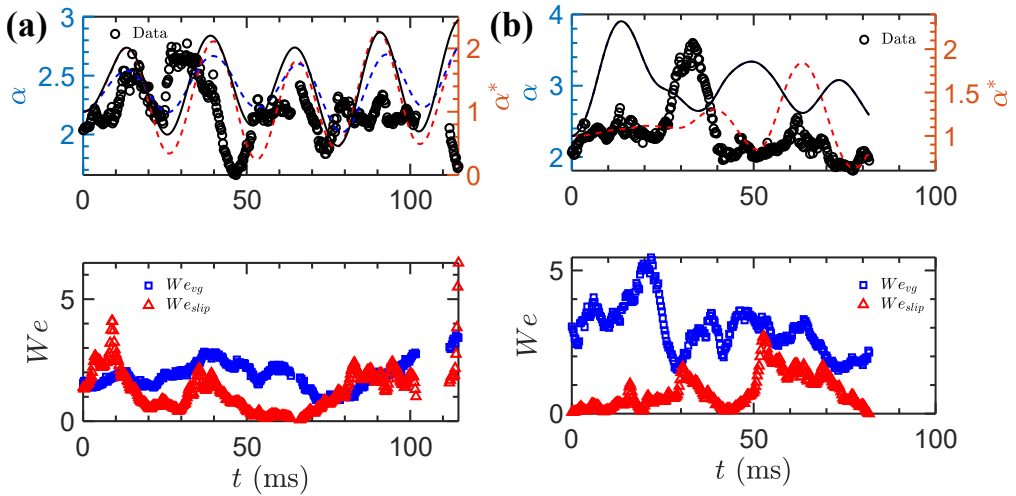}
    \caption{(a-b) Two example time traces of bubble aspect ratio (black circles, top row) and their corresponding Weber numbers (bottom row) during deformation; Each measured Weber number time trace ($We_{vg}$ (blue squares) and $We_{slip}$ (red triangles)) in the bottom panel can be input into equation \ref{eq:risso} to obtain a calculated time trace of $\alpha^*$, which are shown as the dashed lines of the same color in the top panel. In addition, the solid black line represents the calculated results by using max($We_{vg}$,$We_{slip}$) as an input. (Note that the blue dashed line in (b) is not visible because it overlaps perfectly with the black line.)}
    \label{fig:deform_we}
\end{figure}

%
    


Figure \ref{fig:deform_we} shows two examples of the simultaneous measurements of bubble aspect ratio ($\alpha$) and the Weber numbers ($We_{vg}$ and $We_{slip}$). $\alpha$ was calculated as $\alpha=r_1/r_3$ where semi-major axis $r_1$ and semi-minor axis $r_3$ are the instantaneous maximum and minimum radius of a bubble, respectively. These two axes were determined by finding the maximum and minimum vertex-center distances from the 3D-reconstructed bubble geometry, respectively. As shown in figure \ref{fig:deform_we}(a), the reconstruction seems to successfully capture the oscillation of a bubble that undergoes small-amplitude deformation. For this particular case, both of the Weber numbers for almost the entire duration are smaller than 5. The time trace of $\alpha$ is not similar to that of either $We$ at first glance. The only evident correlation is probably at $t=10$--20 ms, when a small peak observed in the time trace of $\alpha$ seems to be driven by a large $We_{slip}$ maybe 5 ms earlier. For $t=50$--70 ms, despite $We_{slip}$ drops close to zero, $\alpha$ continues to rise thanks to a relatively large value of $We_{vg}$ during this period. This indicates that bubbles probably respond to both Weber numbers, likely to be the maximum instantaneous Weber number, i.e. max($We_{vg}$,$We_{slip}$).



The simultaneous measurements also provide a framework to test models for bubble deformation and breakup. One such a model has been proposed before by \citet{risso1998oscillations} and \citet{ lalanne2019model}, which is essentially a forced oscillator model that connects bubble deformation directly to local $We$ through a linear differential equation. This model is designed to follow the interaction between a bubble with surrounding turbulent eddies along its Lagrangian trajectory, exactly how our experiments were performed. The dimensionless form of the equation is:
\begin{equation}
\label{eq:risso}
\frac{d^2\hat{a}}{d\hat{t}^2}+2\xi\frac{d\hat{a}}{d\hat{t}}+\hat{a}=K'We(t)
\end{equation}

where $\xi=1/2\pi\tau_d f_2$ is the damping coefficient, $\tau_d = D^2 / 80 \nu$ is the damping time scale \citep{risso1998oscillations}, and $f_2 = \sqrt{96\sigma / \rho D^3}$ is the mode 2 natural frequency of bubble oscillation \citep{lamb1932hydrodynamics}.

\color{black}
We recognize that this model is linear, and the deformation and breakup process of bubbles in turbulence is nonlinear, especially when bubbles exhibit non-affine deformation and subsequently break. Linearizing this problem relies on two assumptions: (i) The Weber numbers are not very large ($\lesssim O(1)$); and (ii) Bubble deformation is driven primarily by eddies of the bubble size. For (i), similar to the roles played by the Reynolds number in laminar-turbulence transition in single-phase pipe flows, the Weber number here should determine when the process becomes nonlinear. Since there is no consensus on the range of Weber number where the nonlinearity becomes important, we can only argue that the process is linear or nonlinear if the Weber number is much smaller or much larger than $O(1)$. Note that if the flow is turbulent or if the flow Reynolds number is large is irrelevant here. For example, for a small bubble with size smaller than $\eta$ in turbulence, the bubble only senses linear flows around itself even though the flow Reynolds number is large, so the deformation process of this bubble is always linear as long as $We\lesssim 1$. In this work, the Weber number has a wide distribution. For most Hinze-scale bubbles in our experiments with the Weber numbers close to one, the linear model should capture some of the key dynamics. But for highly-deformed bubbles with Weber numbers in the order of $O(10)$ to $O(10^2)$, the linear model is not expected to work, but the discrepancy between the model prediction and measured results may still shed new light on the problem of bubble deformation and breakup in turbulence. For (ii), the Weber numbers defined based on flows of the bubble size essentially ignore the contributions from sub-bubble-scale eddies. Statistically, this assumption probably holds because smaller eddies tend to be weaker, even though they could occasionally become exceedingly strong due to turbulent intermittency. But it would require further investigations to understand their contributions.   

\color{black}

\begin{figure}
    \centering
    \includegraphics[width=0.98\linewidth]{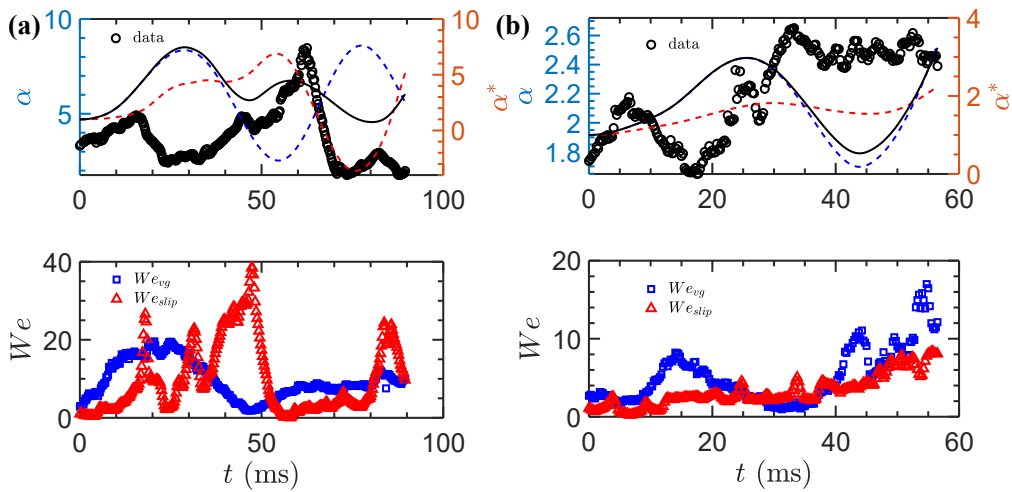}
    \caption{(a-b) Example time traces of two breakup events. Symbols and lines are the same as those in figure \ref{fig:deform_we}. The bubble breaks at 89.5 ms for (a) and at 56.5 ms for (b).}
    \label{fig:breakuptrace}
\end{figure}

In equation \ref{eq:risso}, the amplitude of the instantaneous $We(t)$ is controlled by the prefactor $K'$. $We(t)$ was not available before in other experiments, and it had to be estimated based on two-point velocity measurements from single-phase turbulence \citep{risso1998oscillations}. In this work, in addition to measuring $We(t)$ directly, the method also allows us to distinguish between $We_{vg}$ and $We_{slip}$. But since the model did not explicitly account for individual $We$, here we apply three different inputs: $We(t)=We_{vg}$, $We(t)=We_{slip}$, and $We(t)=\text{max}(We_{vg},We_{slip})$ to equation \ref{eq:risso} to obtain model predictions, which are shown in figure \ref{fig:deform_we} and \ref{fig:breakuptrace} as blue dashed line, red dashed line, and black solid line, respectively. 


Note that $\hat{a}$ in equation \ref{eq:risso} is defined as the ratio of the deformed radius ($a=R-D/2$) to $D$, where $R$ is the major axis of the deformed bubble and $D$ is the diameter of an volume-equivalent sphere. It has to be converted to $\alpha^{*}=2(\hat{a}D+D/2)/D$ for comparisons with the measured $\alpha$. However, despite our best efforts, $\alpha\neq\alpha^{*}$ because $\alpha^{*}$ does not contain information about the minor axis, which has to be replaced with $D/2$. Nevertheless, $\alpha$ and $\alpha^{*}$ should share the same trend as bubbles deform. 

Indeed, similarities can be observed between $\alpha$ and $\alpha^{*}$ in figure \ref{fig:deform_we}(a). The magnitude of $\alpha^{*}$ is affected by the prefactor $K'$ in equation \ref{eq:risso}, which is fixed at 0.1 in this work. In figure \ref{fig:deform_we}(a), the model successfully captures roughly four oscillation periods, which can also be observed in the experimental results. This agreement suggests that the model can explain the deformation and shape oscillations for some bubbles undergoing small-amplitude deformation with small Weber numbers. To be more precise, the Weber numbers, including both $We_{vg}$ and $We_{slip}$, are around 1--3, and the resulting bubble aspect ratio is about 2, which is considered as small-amplitude linear deformation. Note that this definition of small-amplitude deformation and small Weber number are completely based on our observation, and the transition between linear and nonlinear deformation is likely to be smooth over a large range of Weber numbers without having a clean demarcation.

For figure \ref{fig:deform_we}(a), there is a small gap in the time trace with no data at around 100 ms because, during this time period, the velocity gradient calculation does not meet the requirement mentioned in \S\ref{sec:vg}. In addition, despite the overall agreement, the phase lag of each period of $\alpha$ keeps changing in the experimental data, e.g. the second peak is much closer to the first one and further away from the third one. This varying phase lag is a feature that cannot be reproduced from the model. 

Figure \ref{fig:deform_we}(b) shows another example to compare $\alpha$ with $\alpha^*$. A large $\alpha$ observed at 32 ms seems to correlate with an event of large $We_{vg}$ occurred at 20 ms, whereas a small bump of $\alpha$ at 60 ms seems to correspond to a sudden increase of $We_{slip}$ at 50 ms. This observation is still consistent with the argument that bubble deformation tends to be driven by both Weber numbers. In this case, three model-predicted time traces of $\alpha^*$ differ from one another; $We_{vg}$ is systematically larger than $We_{slip}$ for the entire duration. Nevertheless, for this case, although the model-predicted time trace still embraces some oscillatory features, the measured results do not. Over a similar period of time compared with figure \ref{fig:deform_we}(a), only one distinct peak is observed in figure \ref{fig:deform_we}(b). This suggests that the linear-oscillator model provided by \citet{risso1998oscillations} may capture the dynamics of bubbles undergoing small-amplitude  deformation ($\alpha\approx $1--3) with small $We$ ($We\lesssim 3$), like in figure \ref{fig:deform_we}(a), but not for all bubbles, particularly not for bubbles with large $We$ undergoing large aspect ratio changes. In addition, sometimes the surrounding flow maintains its strength, and the bubble is not allowed to oscillate freely. For these cases, the oscillation amplitude becomes smaller, and the model tends to overpredict the bubble aspect ratio. 

Since the model provided by \citet{risso1998oscillations} is primarily designed to characterize the breakup process, figure \ref{fig:breakuptrace} shows two examples of bubbles that eventually break. $We$ for breaking bubbles is clearly much larger: one reaches close to 40 and the other one climbs up to almost 20, nearly a factor of 4--8 larger than the cases for small-amplitude deformation in figure \ref{fig:deform_we}(a). In figure \ref{fig:breakuptrace}(a), $We_{slip}$ dominates, but a local event at $We_{slip}\approx 40$ did not break the bubble, even though it did successfully deform the bubble to a large $\alpha$ at around 8. Following a peak of $We_{slip}$ at $85$ ms, this bubble eventually broke at $t=90$ ms, with the instantaneous $We$ about 10 and local aspect ratio close to 2. 

This example represents many bubbles we observed that do not break at the moment when $\alpha$ reaches its peak; instead, they split at a later time during the process of retraction. As bubbles retract towards a spherical shape, the excess surface energy stored on the bubble interface is transferred back to the surrounding flows in the form of turbulent kinetic energy \citep{dodd2016interaction}. However, this process is unstable because of the large density difference between the two phases, and it eventually leads to breakup before bubbles return back to a spherical shape. 

For the first example (figure \ref{fig:breakuptrace}(a)), $We_{slip}$ is intermittent with a large variation of magnitude in a short period of time, which seems to be consistent with the notion of eddy-bubble collision \citep{prince1990bubble,risso1998oscillations} that this bubble keeps encountering different eddies with varying intensity. For the second example shown in figure \ref{fig:breakuptrace}(b), both Weber numbers slowly increase with time until the bubble breaks. The aspect ratio does not vary much throughout the entire time trace. For the last 20 ms, the aspect ratio of this bubble is close to a constant. Rather than eddy-bubble collision, the results seem to suggest an alternative mechanism: bubbles entrained in an eddy slowly get pulled apart by this eddy as it grows in strength over time.

The model predictions are also shown alongside with these two examples of breakup. In both cases, oscillations clearly observed in $\alpha^*$ from the model calculation are not so obvious in experimental results. As explained before, it is not surprising as large deformation is expected to be highly nonlinear and should deviate from the linear equation \ref{eq:risso}. In addition, this disagreement also implies that the bubble oscillation may not be the right mechanism for bubble breakup at large Weber numbers. Bubbles might just be stretched and deformed by the local strains and slip velocity until the surface tension could no longer hold it together. Such a mechanism that needs to account for 3D couplings between bubbles and surrounding flows is clearly missing in the current model framework. This calls for future investigations into improving the model for large Weber numbers and other breakup mechanisms.

\color{black}


\subsubsection{Distribution of bubble aspect ratio}
In addition to the response of individual bubbles to different Weber numbers, the distribution of $\alpha$ could also be connected to that of $We$ to examine if the bubble aspect ratio can be solely determined by $We$ in a statistical sense. This relationship between $\alpha$ and $We$ was first derived by \citet{moore1965velocity} for bubbles rising in a quiescent medium:
\begin{equation}
\label{eq:moore}
    \alpha=1+\frac{9}{64}\weber+O(We^2)
\end{equation}


where $We$ was defined to account for the dynamic pressure driven by the rising motion of bubbles, not by turbulence. In addition, the key assumption in this model is that $We\ll1$ so that the departure from a spherical shape is so small that any high-order terms associated with $We^2$ can be ignored. 

For $We\approx1$ or above, the potential flow theory applied to oblate ellipsoids with fore-aft symmetry yields:
\begin{equation}
\label{eq:we_alpha}
    We(\alpha) = 4\alpha^{-4/3}(\alpha^3+\alpha-2)[\alpha^2sec^{-1}\alpha-(\alpha^2-1)^{1/2}]^2(\alpha^2-1)^{-3}
\end{equation}

This equation has a maximum aspect ratio of 6 when the Weber number is close to 3.745, above which the symmetric shape is impossible to attain for a bubble. Although this formulation provides a better framework for our studies of bubbles with $\langle We\rangle\approx1$, it cannot predict the relationship between $\alpha$ and instantaneous $We$ for $We>3.745$, which is about 21.5\% of the total events in our experiments. 

\begin{figure}
    \centering
    \includegraphics[width=0.98\linewidth]{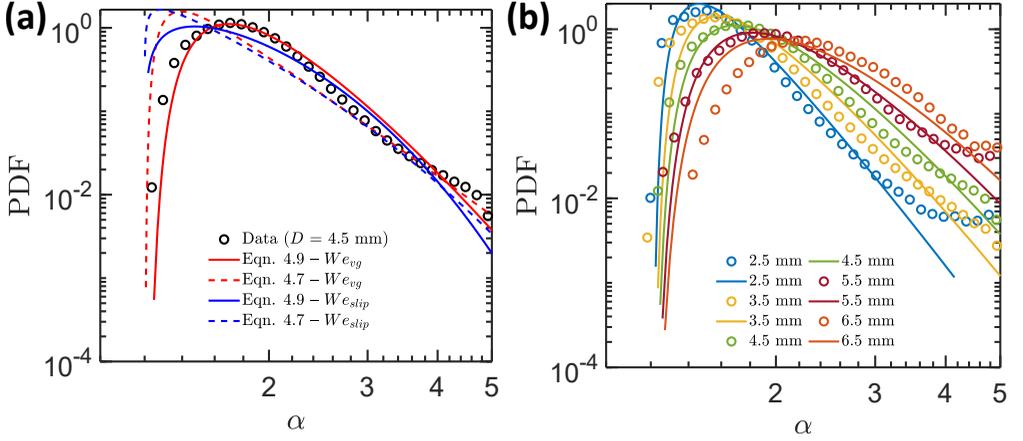}
    \caption{The distribution of bubble aspect ratio $\alpha$ for (a) one size $D=4.5$ mm to test against different $We$ vs. $\alpha$ relationship listed in two different equations \ref{eq:moore} and \ref{eq:we_alpha} by using either $We_{slip}$ (blue) and $We_{vg}$ (red) and for (b) a range of sizes from 2.5 mm to 6.5 mm; solid lines are calculated from equation \ref{eq:aspect_ratio_model}.}
    \label{aspect_ratio_fit}
\end{figure}

\begin{figure}
    \centering
    \includegraphics[width=0.6\linewidth]{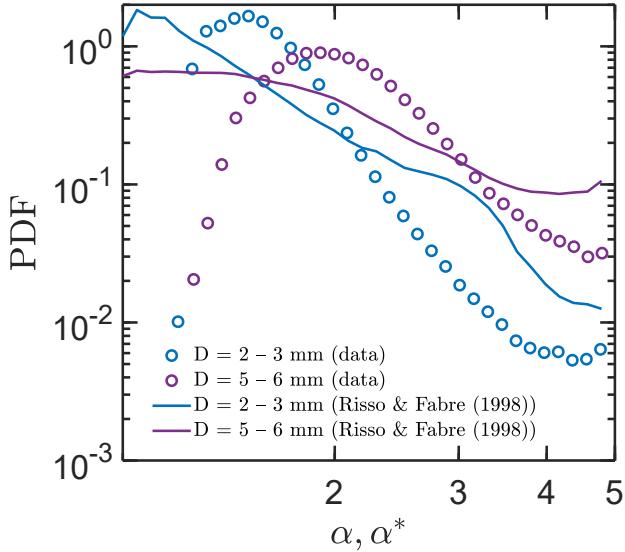}
    \caption{The distribution of bubble aspect ratio $\alpha$ for two different sizes ($D=2$ -- 3 mm and $D=5$ -- 6 mm) from experiments (circle) and linear forced-oscillation model (line, equation \ref{eq:risso}).}
    \label{aspect_ratio_rso}
\end{figure}

Figure \ref{aspect_ratio_fit}(a) shows the PDF of $\alpha$ for bubble size $D=4.5$ mm. The PDF peaks at $\alpha=1.7$ and has a long tail that skews towards larger values of $\alpha$. On top of the experimental results, the PDF of $\alpha$ calculated from equation \ref{eq:moore} using $We=We_{vg}$ as the input is also plotted as the red dashed line. Although the peak location is slightly different from the experimental results, the overall trend is close. Since the long tail in the PDF of $We_{vg}$ comes from the log-normal distribution of $\widetilde{\epsilon}$, a large probability of strong deformation $\alpha>2$ is likely to be contributed by intermittent events with a large $\widetilde{\epsilon}$. If $\alpha$ versus $We_{vg}$ follows a linear relationship, the two PDFs should overlap with each other. To make the red dashed line closer to the experimental results, we tried to add a second-order correction to equation \ref{eq:moore}, which did not provide satisfactory results (not shown here). Finally, after adjusting the parameters in equation \ref{eq:moore}, we settle down to a simple new equation

\begin{equation}
\label{eq:aspect_ratio_model}
    \alpha = \frac{2}{5} We^{\frac{2}{3}} + 1.2
\end{equation}
to fit the data, which is shown as the red solid line in figure \ref{aspect_ratio_fit}(a). This new fit shows an excellent agreement with the measured PDF of $\alpha$. Moreover. the solid blue line in figure \ref{aspect_ratio_fit}(a) shows the PDF of $\alpha$ by implementing a different Weber number $We=We_{slip}$ in equation \ref{eq:aspect_ratio_model}. Similar trend of $\alpha$ can be seen even with this Weber number. But compared with $We_{vg}$, the results based on $We_{slip}$ tend to underpredict $\alpha$, which is consistent with the observation in figure \ref{We_comparison} that the peak of $We_{slip}$ PDF is on the left side of $We_{vg}$ PDF. Nevertheless, the right tail of $\alpha$ can be reproduced by the calculations using both $We_{slip}$ and $We_{vg}$. This may imply that the dynamic stresses contributed by both velocity gradients and the slip velocity are equally important, but turbulent velocity gradients seem to work better and thus more important for mild deformation. 

Figure \ref{aspect_ratio_fit}(b) compiles the PDFs of $\alpha$ for five different bubble sizes, from 2.5 mm to 6.5 mm with an interval of 1 mm. The peaks of these PDFs progressively shift rightward towards a larger $\alpha$ as $D$ grows, which is consistent with our intuition that large bubbles are more deformable and thus have a larger $\alpha$ on average. Moreover, the PDF becomes wider (the right tail of the PDF rises) as $D$ increases, which implies that the probability of bubbles with $\alpha$ much larger than the mean also increases. To explain this, equation \ref{eq:aspect_ratio_model} is applied to all these cases with different $D$, and results are shown as solid lines with corresponding colors to compare with the PDF of measured $\alpha$. The modeled PDF of $\alpha$ agrees with the measured results really well for most bubble sizes except for the largest bubbles where the buoyancy effect may deform bubbles even further. This agreement suggests that the observed change of the PDFs of $\alpha$ as a function of $D$ is driven mostly by the change of $We$, but the relationship between $\alpha$ and $We$ may not be linear for bubbles deformed by turbulence.

An alternative method to predict the relationship between $\alpha$ and $We$ is to use the model provided by equation \ref{eq:risso}. As discussed before, although the model-predicted time trace of $\alpha$ does not match with the measured one exactly, for small Weber numbers, the model is still able to capture some bubble oscillatory deformation, as shown in figure \ref{fig:deform_we}(a). Here, we want to extend the test beyond single time traces. The comparison is shown in figure \ref{aspect_ratio_rso} for only two sizes of 2.5 mm and 5.5 mm for simplicity. The model seems to reproduce the overall trend of the PDF. But the PDF of $\alpha^*$ calculated from the model is more flat than that of the measured $\alpha$, indicating a much higher probability of strongly-deformed bubbles compared with the measured results. This observation is consistent with the time traces shown in figure \ref{fig:breakuptrace} that the linear-oscillator model seems to overpredict the number of large-deformation events. 

This observed difference can also be attributed to other possible reasons. For example, equation \ref{eq:risso} is a linear one-dimensional model with both $We$ and $\alpha^*$ being scalars. In our experiments, $We_{vg}$ has an implicit direction that follows the largest compression direction of the rate-of-strain tensor, and $We_{slip}$ should be aligned with the slip velocity direction. Their combined effect to bubble deformation may not follow a simple relationship of max$(We_{vg},We_{slip})$. In certain circumstances, they could potentially work against each other, which is not accounted for in the linear-oscillator model.  

\color{black}

\subsection{Breakup probability}
\label{sec:breakup_probability}

One condition that is implicitly assumed in many breakup models \citep{martinez1999breakup} is that all bubbles will eventually break, just a matter of time, as long as $We>We_{crit}$. This implies a breakup probability ($\prob$) close to 100\%, which should be valid for large $We\gg1$ and $D\gg D_H$. However, for Hinze-scale or sub-Hinze-scale bubbles with $We\lesssim 1$ and $D\lesssim D_H$, $\prob$ could be anywhere from 0 to 100\%. This number has not been reported before from previous experiments as it is challenging to estimate $\prob$ given the limited residence time of bubbles within the view area. The data that has been generated in this work provides a unique way to evaluate $\prob$ indirectly based on two assumptions: (i) turbulence remains close to homogeneous and isotropic so that the statistics collected in the entire view area can be compiled together to predict the breakup probability; and (ii) the local Weber number is the sole parameter that determines the status of bubble deformation and breakup. Introducing and measuring local $We$ is one step further from the Hinze's seminal work, in which the ensemble-averaged $\langle We\rangle$ was adopted to quantify the breakup probability. The limitation of using $\langle We\rangle$ is that, based on $\langle We\rangle$ being larger or smaller than the critical $We_{crit}$, $\prob$ is close to a step function ($\prob=1$ if $\langle We\rangle>We_{crit}$; $\prob=0$ if $\langle We\rangle<We_{crit}$). However, in turbulence, local flows could be orders of magnitude stronger than the mean; bubbles could break in response to the local $We$ instead of the mean Weber number. To transfer this intuition to quantitative results, the main objective of this section is to determine $\prob$ by linking local $We$ distribution to bubble breakup probability.



\begin{figure}
    \centering
    \includegraphics[width=0.98\linewidth]{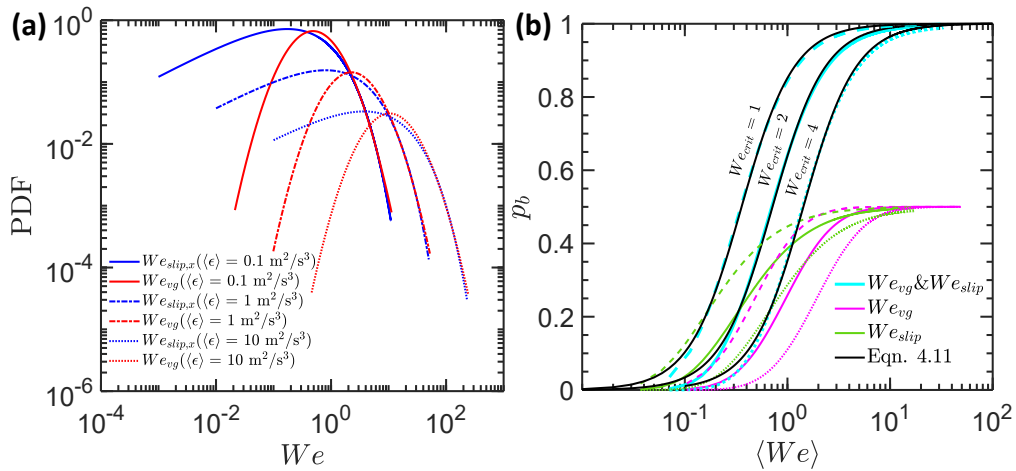}
    \caption{(a) The predicted distribution of both $We_{slip,x}$ and $We_{vg}$ for different energy dissipation rates from $\epsilon$=0.1 to 10 m$^2$/s$^3$. (b) Breakup probability $p_b$ calculated based on different mean Weber number $\langle We\rangle$. Three sets of lines indicate three different $We_{crit}$ from 1 to 4. Within each set, $p_b$ based on the total Weber number (cyan), or either $We_{vg}$ (red) or $We_{slip}$ (blue) alone, are shown. The curves from the total $We$ were fitted with equation \ref{eq:p_b} to predict $p_b$ for any mean $We$ and any selected $We_{crit}$. }
    
    \label{fig:we_predict}
\end{figure}

Before estimating $p_b$, we would like to extend the PDF of $We_{vg}$ and $We_{slip}$ beyond our experiments to other turbulent flows with different $\langle\epsilon\rangle$ and bubble size $D$. Based on equations \ref{eq:intm} and \ref{eq:slip}, the distribution of local $We$ for three $\langle\epsilon\rangle$ from 0.1 m$^2$/s$^3$ to 10 m$^2$/s$^3$ are shown in figure \ref{fig:we_predict}(a). Both $We_{slip}$ and $We_{vg}$ shift rightward systematically with a seemingly-identical shape. It is important to note that the shape remains the same on the logarithmic scale, indicating that the distribution actually widens on the linear scale and the standard deviation of local $We$ increases as $\langle\epsilon\rangle$ grows. 

From the distribution, we can estimate $p_b$ based on:
\begin{equation}
p_b(\langle We\rangle)=\int^{+\infty}_{We_{crit}}p(We) d(We)    
\end{equation}


where the local $We$ could be either $We_{vg}$ or $We_{slip}$, and $\langle We\rangle=\int^{+\infty}_{-\infty}[We\times p(We)] d(We)    
$. Here $p_b$ is written as a function of the mean Weber number $\langle We\rangle$ to help with other experiments or simulations that have access only to $\langle We\rangle$ from the mean energy dissipation rate. It can be seen that, as $\langle We\rangle$ grows, either due to a larger $D$ or larger $\epsilon$, the bubble breakup probability $p_b$ will increase. 


Figure \ref{fig:we_predict}(b) shows $p_b$ as a function of $\langle We\rangle$. In most previous works assuming that the breakup probability has a sharp transition at $We_{crit}$, $p_b$ would behave like a step function: $p_b=1$ for $\langle We \rangle\geq We_{crit}$ and $p_b=0$ for $\langle We \rangle< We_{crit}$. Although these two limits still apply in figure \ref{fig:we_predict}(b), the transition is much smoother, spanning over a few orders of magnitude of $\langle We \rangle$. Note that the choice of $We_{crit}$ does not affect the shape of the curve. As shown in figure \ref{fig:we_predict}(b), when we change $We_{crit}$ from 1 to 4, it just shifts the transitional $\langle We\rangle$ towards the new $We_{crit}$ without affecting the overall trend. 

This framework applies to $We_{vg}$ and $We_{slip}$, both of which contribute to bubble breakup. Since the right tail of their respective distribution is very close to each other, an equal contribution from the two breakup mechanisms was assumed. As the result, the curves of $p_b$ for either $We$ alone approach 0.5 for $\langle We\rangle$ much larger than $We_{crit}$ to ensure that the sum of the two $p_b$ equals to one. Furthermore, even for the same $We_{crit}$, $p_b$ of $We_{slip}$ (blue lines) stays mostly above that of $We_{vg}$ (red lines) until they cross at a location very close to the plateau near $p_b=$0.5. This difference can be ascribed to the difference of the PDFs: compared with $We_{vg}$, the PDF of $We_{slip}$ has a larger probability for small $We$. This also suggests that bubbles close to the Hinze scale may be deformed more often by the slip velocity.



Finally, the total breakup probability $p_b$ by summing the contribution from $We_{vg}$ and $We_{slip}$ and using $We_{crit}=1$ to 4 are shown as three cyan  lines in figure \ref{fig:we_predict}. These three lines are fitted with the same switch function:
\begin{equation}
\label{eq:p_b}
    p_b=[1 + (2.8\langle We\rangle/We_{crit})^{-1.7}]^{-1}
\end{equation}
which includes $We_{crit}$ as the input. The fitted results are shown in figure \ref{fig:we_predict}(b) as three black solid lines. One may not see the cyan lines at all because the fit overlap perfectly with the calculated results over the entire $\langle We\rangle$ range for three $We_{crit}$ considered. Equation \ref{eq:p_b} provides a method to estimate bubble breakup probability in turbulence, particularly for bubbles close to the Hinze scale and $\langle We \rangle\approx 1$. 

\section{Conclusion}\label{conclusion}

Bubble deformation and breakup in intense turbulence is ubiquitous in many applications, but details of how this takes place for a bubble close to the Hinze scale remain elusive because of the lack of data to probe the interaction between finite-sized bubbles and surrounding turbulence. In this study, both 3D bubble geometry and nearby 3D particle tracks were acquired simultaneously using our in-house virtual camera reconstruction and particle tracking algorithm. The experiments were performed in a system that can reach a high turbulent energy dissipation rate that can significantly deform and even break bubbles, while maintaining homogeneous and isotropic turbulence throughout the entire measurement volume. 
Since the 3D information of both phases is available, this unique dataset allows us to interrogate the couplings between the two phases, in particular the key mechanisms that drive bubble deformation and breakup. The flow velocity was decomposed into two components, the local flow velocity and velocity gradient, both coarse-grained at the bubble scale. Each component can be used to define its own Weber number as a way to quantify their relative contributions to bubble deformation. 

In this study, in addition to directly measuring the Weber numbers, bubble deformation is also connected to the log-normal distribution of the local coarse-grained energy dissipation rate $\epsr$. The modeled distribution of both $\epsr$ and $We_{vg}$ based on turbulence characteristics agree well with the measured results. Moreover, because of the density mismatch between the two phases and the finite bubble size effect, the slip velocity also plays an important role. Based on this observation, a different Weber number is defined to measure deformation driven by the slip velocity, whose distribution can be fitted with a stretched exponential function inspired by the distribution of two-point velocity increments in single-phase turbulence. Based on this function, the distribution of the slip-velocity-based Weber number can be connected to $\langle\epsilon\rangle$ and bubble size.  

The distribution of the Weber number was also connected to the reconstructed bubble geometry. It has been shown that the relationship that was developed for describing bubbles rising in a quiescent medium does not work well for the turbulent case. A new non-linear model was proposed to improve the fit, and it seems to work well for a range of bubble sizes considered. In addition, the results were tested against a linear forced oscillator model that was proposed before. Although the model does seem to reproduce some key features of a few example time traces qualitatively, the distribution of the predicted aspect ratio does not match with the directly-measured results quantitatively. 

Finally, the Weber number distribution is generalized for different bubble sizes and energy dissipation rates in order to evaluate breakup probability, which was estimated based on the mean energy dissipation rate in many other works. In contrast to what has been proposed before that the bubble breakup probability experiences a precipitous drop as bubble size decreases below the Hinze scale, accounting for the distribution of local Weber number helps to smooth the curve near the Hinze scale. The final calculated relationship between breakup probability and the mean Weber numbers was fitted with a simple function that can help future works to estimate bubble breakup probability based on the mean Weber number.

We acknowledge the financial
support from the National Science Foundation award numbers: 1854475 and CAREER-1905103. We would also like to acknowledge Charles Meneveau for suggestions. Declaration of Interests: The authors report no conflict of interest.

\bibliographystyle{jfm}
\bibliography{ref}

\end{document}